\documentclass[superscriptaddress,nofootinbib,showkeys,showpacs,byrevtex,preprintnumbers]{revtex4}
\usepackage{graphicx}
\usepackage{epstopdf}
\usepackage{MnSymbol}
\begin{document}
\preprint{PKNU-NuHaTh-2016-01}
\preprint{CYCU-HEP-16-02}
\title{QCD topological susceptibility from the nonlocal chiral quark model}
\author{Seung-il Nam}
\email[E-mail: ]{sinam@pknu.ac.kr}
\affiliation{Department of Physics, Pukyong National University (PKNU), Busan 608-737,
Republic of Korea}
\affiliation{Asia Pacific Center for Theoretical Physics (APCTP), Pohang 790-784,
Republic of Korea}
\author{Chung-Wen Kao}
\email[E-mail: ]{cwkao@cycu.edu.tw}
\affiliation{Department of Physics and Center for High Energy Physics, Chung-Yuan Christian University, Chung-Li 32023, Taiwan}
\date{\today}
\begin{abstract}
We investigate the QCD topological susceptibility $\chi_t$ by using the nonlocal chiral quark model (NL$\chi$QM). This model is based on the liquid instanton QCD-vacuum configuration in which $\mathrm{SU}(3)$ flavor symmetry is explicitly broken by the current quark mass $(m_{u,d},m_s)\approx(5,135)$ MeV. To compute $\chi_t$, the local topological charge density operator $Q_t(x)$ is derived from the effective partition function of NL$\chi$QM. We take into account the contributions from the leading-order (LO) ones $\sim\mathcal{O}(N_c)$ in the $1/N_c$ expansion. We also verify that the analytical expression of $\chi_t$ in NL$\chi$QM satisfy the Witten-Veneziano (WV) and the Leutwyler-Smilga (LS) formulae. Once the average instanton size and inter-instanton distance are fixed with $\bar{\rho}=1/3$ fm and $\bar{R}=1$ fm, respectively, all the associated model parameters are all determined self-consistently within the model, including the $\eta$ and $\eta'$ weak decay constants. We obtain the results such as $F_{\eta}=96.77$ MeV and $F_{\eta'}=102.53$ MeV for instance. Numerically we observe that $\chi_{t}=(165.57\,\mathrm{MeV})^4$ in our full calculation. This value is comparable with its empirical one $\chi_t=(175\pm5\,\mathrm{MeV})^4$. We also find that our $\chi^\mathrm{WV}_t=\chi^\mathrm{QL}_t=(194.30,\mathrm{MeV})^4$ in the quenched limit and $\chi^\mathrm{LS}_t=(162.54\,\mathrm{MeV})^4$ in the chiral limit. Consequently, we conclude that $\chi_{t}<\chi^\mathrm{QL}_t$. Our result also implies that the $(10\sim20)\,\%$ decrease with the dynamical quark contributions.
\end{abstract}
\maketitle
\section{introduction}
The structure of vacuum of Quantum Chromodynamics (QCD) is one of the most important subject in hadronic physics and has been intensively studied for decades. In particular, it is impossible to understand the phase structure of QCD and the hadron spectrum without good understanding of the QCD vacuum.
For example, it has been well known that the spontaneous chiral symmetry breaking (S$\chi$SB) of the QCD vacuum plays the crucial role in the QCD phenomenology in the low energy regime.
The most important consequence of S$\chi$SB is the existence of the massless modes generated
via the Nambu-Goldstone mechanism identified as the pseudo-scalar (PS) mesons in the hadron spectrum.
Furthermore their masses can be explained by the small but finite current quark masses $m_{u,d}\sim5$ MeV which break the chiral symmetry explicitly. The kaons are heavier than the pions because that the $\mathrm{SU}(3)$ flavor symmetry is explicitly broken by the much heavier strange quark mass $m_s\sim135$ MeV. S$\chi$SB also explains the absence of the mass degeneracy between the parity partners such as $N(940,\frac{1}{2}^{+})$ and $N(1535,\frac{1}{2}^{-})$ which is a consequence of the chiral symmetry of QCD Hamiltonian. Moreover one can construct the chiral effective Lagrangian to calculate the interactions between hadrons in the low energy regime.

Although most of the mass spectrum of the light PS mesons can be understood by the spontaneous and explicit chiral symmetry breaking, the large mass splitting between the $\eta$ and $\eta'$ mesons in the flavor $\mathrm{U}(3)$ nonet had remained a puzzle. Even including the current quark masses effect, the mass of  $\eta'$ should satisfy $m_{\eta'}\le \sqrt{3}m_{\pi}$, if $\mathrm{U_A}(1)$ is spontaneously broken~\cite{Weinberg:1975ui}. This puzzle was later resolved by the $\mathrm{U_A}(1)$ anomaly~\cite{Weinberg:1975ui}, namely $\mathrm{U_A}(1)$ symmetry is already broken at quantum level. 'tHooft first suggested that the {\it instanton} breaks the $\mathrm{U_A}(1)$ symmetry explicitly~\cite{'tHooft:1976up} and constructed a dilute instanton gas model which is a semi-classic approximation of the topological charge distribution of Yang-Mills theory to explain $\mathrm{U_A}(1)$ problem. However the approximation itself breakdowns in the infrared regime such that it cannot provide the estimation of the $\eta'$ mass. The strength of this anomaly is characterized by the topological susceptibility $\chi_t$ defined by the following two-point correlation at zero-momentum transfer:
\begin{equation}
\label{eq:TS}
\chi_t\equiv\lim_{V\to\infty}\frac{\langle Q^2_t\rangle}{V}
=\int_x\,\langle 0|T\left(Q_t(x)Q_t(0)\right)|0\rangle,
\end{equation}
where $Q_t(x)$ indicates the topological charge density operator and $V$ stands for the four-dimensional volume in Euclidean space. 'tHooft first suggested that the {\it instanton} breaks the $\mathrm{U_A}(1)$ symmetry explicitly~\cite{'tHooft:1976up} and constructed a dilute instanton gas model which is a semi-classic approximation of the topological charge distribution of Yang-Mills theory to explain $\mathrm{U_A}(1)$ problem. However the approximation itself breakdowns in the infrared regime such that he cannot provide the estimate of the $\eta'$ mass. On the other hand, the first estimation of $\eta'$ mass was given by the Witten-Veneziano formula in the pure Yang-Mills (YM) QCD in the large $N_c$ limit. It is derived from the Ward identity of the flavor singlet chiral current~\cite{Veneziano:1979ec,Witten:1979vv}:
\begin{equation}
\label{eq:CHPT}
\chi_t=\frac{F^2_\pi}{2N_f}\left(m^2_\eta+m^2_{\eta'}-2m^2_K\right).
\end{equation}
Empirically, using Eq.~(\ref{eq:CHPT}) by plugging in the physical masses of the PS mesons one has $\chi_t\approx(175\pm5\,\mathrm{MeV})^4$~\cite{Veneziano:1979ec}. This value has been confirmed by the lattice simulation~\cite{Alles:1996nm}. Furthermore, the dilute gas model of 'tHooft and Witten-Veneziano approach differ from each other in the ratio $\langle Q^4_t\rangle/\langle Q^2_t\rangle$ and the numerical result seems to prefer the latter.~\cite{Ce:2014sfa}. Since the topological susceptibility $\chi_t$ plays such an important role in QCD, naturally there have been many theoretical approaches to compute $\chi_t$ including many lattice QCD (LQCD) simulations~\cite{DelDebbio:2004ns,AliKhan:1999zi,AliKhan:1999zp,Hart:2000wr,Hart:2000gh,Bali:2001gk,Bernard:2012fw,Chiu:2008jq,Aoki:2000kp}, effective QCD-like models~\cite{Fukushima:2001hr,Xia:2013caa,Cheng:1998vj,Dorokhov:2004pw}, QCD sum rule~\cite{Grunberg:1984pa}, $\chi_t$ at finite temperature~\cite{Jiang:2015xqz,Alles:1996nm,Borsanyi:2015cka}, AdS/QCD approach~\cite{Katz:2007tf}, and so on.

In the present work, we employ the nonlocal chiral quark model (NL$\chi$QM) to compute $\chi_t$.
The present model is based on the liquid instanton QCD-vacuum configuration. After the bosonization, the effective partition function of the instanton model becomes a partition function with quarks ($q$) and pseudo-scalar (PS) mesons ($\phi$) with the nonlocal interactions between them. Therefore it is called the nonlocal chiral quark model (NL$\chi$QM,). To identify the topological charge operator $Q_t(x)$ within NL$\chi$QM, we need calculate the four divergence of the flavor singlet current derived by the infinitesimal $\mathrm{U_A}(1)$ rotation of the partition function. From it one can figure out the topological charge operator. 

We also consider the small but finite difference $(\Delta)$ between the numbers of the instantons $(N_{+})$ and the anti-instantons $(N_{-})$ in order to derive the appropriate expression of $\chi_t$ in the chiral limit. By computing the two-point correlation defined in Eq.~(\ref{eq:TS}) by using $Q_t(x)$ of our own model, we obtain basically three distinctive contributions for $\chi_t$ for the leading order (LO) in the $1/N_c$ expansion:
1) The contribution with the dynamical quarks $\chi^q_t$, which contains the quark loops and the PS-meson propagator.
2) That consists of the PS-meson propagator without the quark loops $\chi^\phi_t$.
3) A constant contribution from the $\Delta$ effect, i.e. $\chi^\Delta_t$, which survives only in the chiral limit $m_q\sim0$ where $m_q$ stands for the current quark mass with its flavor $q=(u,d,s)$. Furthermore, we verify that NL$\chi$QM reproduces the Witten-Veneziano and Leutwyler-Smilga formulae analytically.

After fixing the average (anti)instanton size $\bar{\rho}=1/3$ fm and inter-(anti)instanton distance $\bar{R}=1$ fm, we generate all the physical quantities needed for computing $\chi_t$ self-consistently in NL$\chi$QM. Our values of the masses and weak decay constants of the PS mesons are in a good agreement with the experimental values. Our numerical value of $\chi_{t}=(165.57\,\mathrm{MeV})^4$ whereas $\chi^\mathrm{QL}_t=(194.30\,\mathrm{MeV})^4$ in the quenched limit. In other words, we find that the inclusion of the dynamical quarks reduces $\chi_t$ by $(10\sim20)\%$ in NL$\chi$QM. It is consistent with the LQCD data~\cite{Bali:2001gk}. Our result is also comparable with other theoretical results and the empirical value $\chi_{t}=(175\pm5\,\mathrm{MeV})^4$.

The present work is organized as follows: In Section II, we briefly introduce our theoretical framework including the details of NL$\chi$QM and the derivation of the expression of $\chi_t$ within NL$\chi$QM. Our numerical results and related discussions are given in Section III. The last Section is devoted to the conclusion and future perspective.
 \section{Theoretical framework}
The local topological charge density operator $Q_t(x)$ of QCD is defined by~\footnote{In literatures, the topological charge is given by $Q_t$, and its density operator by $q_t(x)$. However, in the present work, we will make use of $Q_t(x)$ as the density operator.}
\begin{equation}
\label{eq:TC}
Q_t(x)\equiv \frac{1}{32\pi^2}G^a_{\mu\nu}(x)\tilde{G}^a_{\mu\nu}(x),\,\,\,\,
\tilde{G}^a_{\mu\nu}(x)=\frac{1}{2}\epsilon_{\mu\nu\sigma\rho}G^a_{\sigma\rho}(x).
\end{equation}
Here, $G^a_{\mu\nu}$ and $\tilde{G}^a_{\mu\nu}$ denote the gluon field-strength tensor and its dual one, respectively.
Note that the strong coupling $g_s$ is implicitly included in $G^a_{\mu\nu}$ in Eq.~(\ref{eq:TC}). The topological charge of QCD can be obtained by integrating $(Q_t(x))^2$ over the volume.
A non-zero topological charge implies that the tunneling occurs between the QCD vacua related to each other by the homotopy of the SU(3) color-gauge symmetry. It is worth mentioning that the {\it instanton} is one of the semi-classical solutions to explain this novel QCD phenomenon~\cite{'tHooft:1976up}. Moreover, a non-zero topological charge also indicates that the flavor-singlet axial current of QCD is not conserved because of the $\mathrm{U_A}(1)$ anomaly. Thus, the instanton can explain the $\mathrm{U_A}(1)$ symmetry breaking explicitly~\cite{Diakonov:1995qy}. In QCD with the vacuum angle $\theta\to0$, the four divergence for the $\mathrm{U_A}(1)$ current reads in Euclidean space:
\begin{equation}
\label{eq:CSC}
\partial^\mu J^\mathrm{singlet}_{5\mu}(x)=2N_fQ_t(x)+2\sum_{q=u,d,s}m_qq^\dagger(x)\gamma_5 q(x),
\end{equation}
where $q(x)$ and $m_q$ denote the field and the current mass for a quark. $N_f$ is the number of the flavors of quarks.
To obtain $Q_t(x)$ from an effective model of QCD, one can apply Eq.~(\ref{eq:CSC}) to obtain $\partial^\mu J^\mathrm{singlet}_{5\mu}(x)$ by differentiating the effective Lagrangian of the model with respect to the infinitesimal rotation parameter $\epsilon$ as in $q\to\left(1+i\gamma_5\frac{\epsilon}{2}\right)q$:
\begin{equation}
\label{eq:DL}
\frac{\partial(\delta\mathcal{L}_\mathrm{eff})}{\partial\epsilon}
\equiv \partial^\mu J^\mathrm{singlet}_{5\mu}.
\end{equation}
Comparing Eq.~(\ref{eq:CSC}) with Eq.~(\ref{eq:DL}), one is able to obtain the expression of $Q_t(x)$ and compute the topological susceptibility.
\begin{figure}[t]
\includegraphics[width=14cm]{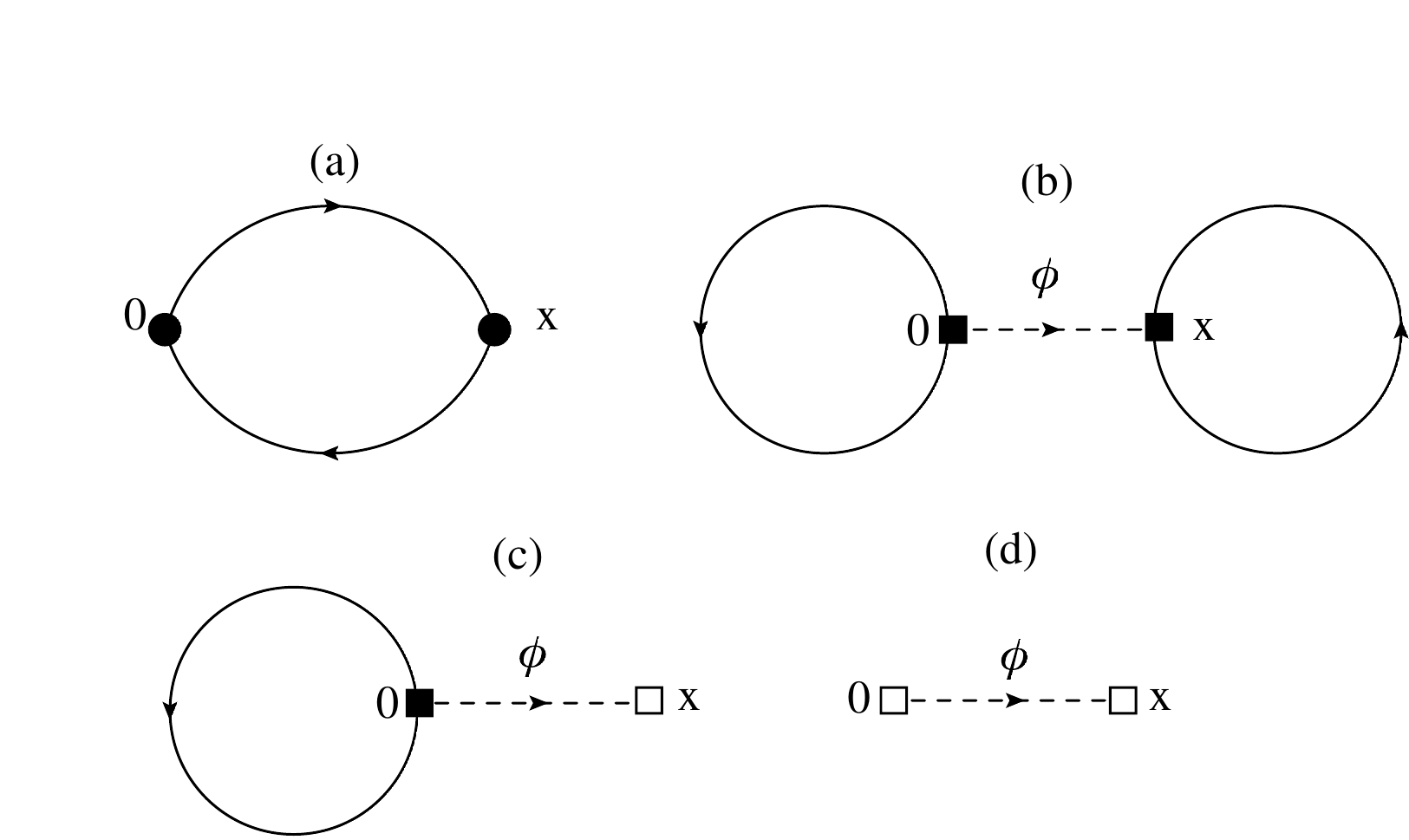}
\caption{The leading-order diagrams of the topological susceptibility $\chi_t$ in NL$\chi$QM in the large $N_c$ limit. The solid and dash line indicates the quark and pseudo-scalar meson, whereas the solid ($\blacksquare$) and blank ($\square$) squares represent the vertices in $\mathcal{O}(1/\sqrt{N_c})$ and $\mathcal{O}(\sqrt{N_c})$, respectively, and the solid circle ($\bullet$) stands for the vertex in $\mathcal{O}(1)$.}
\label{FIG0}
\end{figure}

In this article, we use the partition function derived from the dilute liquid-instanton QCD-vacuum model (LIM)~\cite{Diakonov:1997sj,Diakonov:1995qy,Diakonov:2002fq} in Euclidean space:
\begin{eqnarray}
\label{eq:EZ}
\mathcal{Z}_\mathrm{eff}[q,q^\dagger]&=&\int\frac{d\lambda_\pm}{2\pi}\int Dq Dq^\dagger
\exp\left[
\int_x\sum_{q}q^\dagger(i\rlap{/}{\partial}+m_q)q+\sum_{a=\pm}\left[\lambda_a Y^a_{N_f}(\bar{\rho})+N_a\left(\ln\frac{N_a}{\lambda_a V \mathrm{\Lambda}^{N_f}}\right)\right]\right],
\cr
Y^a_{N_f}(\bar{\rho})&=&\frac{1}{N^{N_f}_c}\int_x \,
\mathrm{det}_f\left[iJ^a_{qq'}(x,\bar{\rho})\right]=
\int_x\,\mathrm{det}_f\left[\frac{i}{N_c}J^a_{qq'}(x,\bar{\rho})\right],
\cr
J^a_{qq'}(x,\bar{\rho})&=&\int_k\int_pe^{i(k-p)\cdot x}F(k)F(p)\left[q^\dagger(k)\frac{1+a\gamma_5}{2}q'(p)\right]_{N_f\times N_f}.
\end{eqnarray}
Here, we assign $\int\frac{d^4k}{(2\pi)^4}$ with $\int_k$ for brevity, while $[\cdots]_{N_f\times N_f}$ and $\mathrm{det}_f$ denote the $(N_f\times N_f)$ quark-flavor matrix and the determinant over the flavor indices. $\hat{m}_q$ indicates the current-quark mass matrix. In the case of $N_f=3$ one has
$\mathrm{diag}(m_u,m_d,m_s)$. The parameter $a$ indicates the instanton ($+$) and anti-instanton ($-$) contributions. Thus, $N_a$ and $\lambda_a$ denote the number of (anti)instantons and the Lagrange multiplier, respectively. $V$ stands for the four-dimensional volume in Euclidean space, whereas $\mathrm{\Lambda}$ for an argument to make the logarithm dimensionless.

In deriving Eq.~(\ref{eq:EZ}), we have assumed the instanton distribution function to be a $\delta$ function, $\delta(\rho-\bar{\rho})$. Here $\rho$ stands for the (anti)instanton size and $\bar{\rho}$ their average value. $F(k)$ stands for the Fourier transform of the instanton zero-mode solution. It plays the role of UV regulator, i.e. form factor (FF) and its explicit form will be given in the next Section.
Furthermore, we assume that $N_+\approx N_{-}\approx N/2=(N_++N_-)/2$ and $\lambda_{+}\approx \lambda_{-}=\lambda$. However, we will consider the small number difference between the instanton and anti-instanton $\Delta=|N_+-N_-|\ll N$ later to interpret the correct behaviour of $\chi_t$ in the chiral limit. Note that the determinant-type interaction, i.e. $2N_f$ 't Hooft interaction is derived directly from the instanton configuration. It gives the explicit $\mathrm{U_A}(1)$ symmetry breaking.

In order to obtain $Q_t(x)$ from Eq.~(\ref{eq:EZ}), we perform the infinitesimal chiral rotation as shown in Eq.~(\ref{eq:DL}) which results in $\mathcal{Z}_\mathrm{eff}\to\mathcal{Z}'_\mathrm{eff}$:
\begin{eqnarray}
\label{eq:EFFPART}
\mathcal{Z}'_\mathrm{eff}[q,q^\dagger]&=&\int\frac{d\lambda_\pm}{2\pi}\int Dq Dq^\dagger
\exp\left[
\int_x\sum_{q}q^\dagger(i\rlap{/}{\partial}+im_q-2\epsilon\gamma_5m_q)q
+\sum_{a=\pm}\left[\lambda_a Y'^a_{N_f}(\bar{\rho})+N_a\left(\ln\frac{N_a}{\lambda_a V \mathrm{\Lambda}^{N_f}}\right)\right]\right],
\cr
Y'^a_{N_f}(\bar{\rho})&=&
(1+2aiN_f\epsilon)\int_x\,\mathrm{det}_{N_f}\left[\frac{i}{N_c}J^a_{qq'}(x,\bar{\rho})\right].
\end{eqnarray}
Note that there are additional terms proportional to $\epsilon$ which break the $\mathrm{U_A}(1)$ symmetry. If we put $F(k)=1$ then perform the Fourier transform of the quark fields, it is easy to find that those additional terms in Eq. (\ref{eq:EFFPART}) coincide with the NJL model result of Eq.~(13) in Ref.~\cite{Fukushima:2001hr}. There they tried to include the {\it ring} diagrams in addition to the leading contributions in $\mathcal{O}(N_c)$
in the large $N_c$ limit in terms of the mean-field approach. Those ring diagrams were identified as the PS-meson propagators.
Here we will not follow their strategy. Instead, we prefer to apply the technique of bosonization to
convert the partition function in Eq.~(\ref{eq:EFFPART}) into the one with quarks and
the auxiliary meson fields $\mathcal{M}$ in the large $N_c$ limit~\cite{Diakonov:1997sj}.

The first step is to employ the following identity for the bosonization:
\begin{equation}
\label{eq:B}
\exp\left(\lambda_{\pm}\mathrm{det}_f\left[\frac{i}{N_c}J^\pm_{qq'}\right] \right)=\int D\mathcal{M}_{\pm}\exp\left\{
i\mathrm{tr}_f\left[\mathcal{M}^{qq'}_\pm J^\pm_{qq'}\right]
-(N_f-1)\left(\frac{1}{\lambda_{\pm}}\mathrm{det}_f[\mathcal{M}^{qq'}_\pm N_c] \right)^{\frac{1}{N_f-1}}\right\},
\end{equation}
where $\mathrm{tr}_f$ stands for the trace over the flavor indices.
$\mathcal{M}_{\pm}$ stands for the left-handed $(+)$ and right-handed $(-)$ nonlinear auxiliary fields. Since, in the bosonized partition function, the interactions between the quarks and mesons are nonlocal, and the chiral symmetry is broken dynamically, therefore it is dubbed the nonlocal chiral-quark model (NL$\chi$QM). After performing the bosonization, the rotated effective partition function in Eq.~(\ref{eq:EFFPART}) becomes
\begin{eqnarray}
\label{eq:EFFPART1}
&&\mathcal{Z}'_\mathrm{eff}[q^\dagger,q,\mathcal{M}_{\pm}]=
\int\frac{d\lambda_\pm}{2\pi}\int Dq Dq^\dagger D\mathcal{M}_\pm
\exp\Bigg\{
\int_x\sum_qq^\dagger(i\rlap{/}{\partial}+i\hat{m})q
+\sum_{a=\pm}N_a\left(\ln\frac{N_a}{\lambda_a V \mathrm{\Lambda}^{N_f}}\right)
\cr
&+&\sum_{a=\pm}\int_x\Bigg[i\mathrm{tr}_f[\mathcal{M}^{qq'}_a J^a_{qq'}]
-(N_f-1)\left(\frac{1}{\lambda_a}\mathrm{det}_f[\mathcal{M}^{qq'}_a N_c]
 \right)^{\frac{1}{N_f-1}}
\Bigg]
\cr
&+&\epsilon\left[2i\int_x\sum_qq^\dagger(i\gamma_5m_q)q+ 2iN_f\sum_{a=\pm}a \int_x\,\Bigg[i\mathrm{tr}_f[\mathcal{M}^{qq'}_a J^a_{qq'}]-(N_f-1)\left(\frac{1}{\lambda_a}\mathrm{det}_f[\mathcal{M}^{qq'}_a N_c] \right)^{\frac{1}{N_f-1}}
\Bigg]\right]\Bigg\}.
\end{eqnarray}

Focusing on the deviations by the infinitesimal rotation in $\epsilon[\cdots]$ and comparing them with Eq.~(\ref{eq:CSC}),
one can define the local topological charge density operator from NL$\chi$QM:
\begin{equation}
\label{eq:TCO}
Q_t[q_f,q^\dagger_f,\mathcal{M}_{\pm}]= \sum_{a=\pm}a \int_x\,
\left[\mathrm{tr}_{f}[\mathcal{M}^{qq'}_a J^a_{qq'}]+i(N_f-1)\left(\frac{N_c^{N_f}}{\lambda_a}\mathrm{det}_{f}\left[\mathcal{M}^{qq'}_a\right] \right)^{\frac{1}{N_f-1}}
\right].
\end{equation}
Following Refs.~\cite{Diakonov:1997sj,Halperin:1998gx} we parameterize the auxiliary meson field $\mathcal{M}^{qq'}_a$.
\begin{equation}
\label{eq:PSP}
\mathcal{M}^{qq'}_\pm\to\sigma_0(1\pm\delta)(U_\pm)_{qq'}=\sigma_0(1\pm\delta)(e^{\pm i\Phi})_{qq'}.
\end{equation}
Here $\sigma_0$ stands for VEV of the flavor-singlet scalar meson, and plays the role to provide the quarks with their
constituent quark masses. In other words, one has the constituent quark mass $M_0=\sigma_0$.
Furthermore we have $M_q=M_0+m_q$ since the quark mass is the sum of the current quark mass and the constituent quark mass.
To determine the value of $\sigma_0$, one minimize the effective action with respect to $\lambda$. By this way, one obtains
\begin{equation}
\label{eq:NOV}
\frac{N}{V}=4N_c\int_k\frac{M^2_0(k)}{k^2+M^2_0(k)},\,\,\,\,M_0(k)=\sigma_0 F^2(k),\,\,\,\,
\lambda=\frac{N}{2V}\left(\frac{2\sigma_0 V N_{c}}{N}\right)^{N_f}.
\end{equation}

We also introduced a new parameter $\delta$ in ~Eq.~(\ref{eq:PSP}) defined as follows~\cite{Diakonov:1995qy}:
\begin{equation}
\label{eq:DELTA}
\delta=\frac{2\pi^2\bar{\rho}^2\Delta}{VN_c}\left(\sum_q\frac{1}{m_qM_q}\right)
=\frac{\Delta}{V}\left(\sum_q\frac{1}{m_q\Sigma_q}\right)
=Q^\mathrm{YM}_t\left(\sum_q\frac{1}{m_q\Sigma_q}\right).
\end{equation}
where $\Delta=|N_+-N_-|\ll N$ is the small difference between the numbers of instantons and anti-instantons.
$Q^\mathrm{YM}_t=\frac{\Delta}{V}$ is the topological charge operator in pure Yang-Millers theory.
Note that $\delta$ is a quantity of $\mathcal{O}(1/N_{c})$.
$\Sigma_q$ denotes the quark condensate for the flavor $q$  which is related with $M_q$ by the following relation:
\begin{equation}
\Sigma_q=M_q\left(\frac{N_c}{2\pi^2\bar{\rho}^2}\right).
\end{equation}
Furthermore, $U_\pm$ can be understood as the $N_f\times N_f$ nonlinear PS-meson field.
Since we are interested in the contributions from the iso-singlet PS mesons for the $\mathrm{U_A}(1)$ anomaly, 
we pick up only $\eta$ and $\eta'$ from the flavor $\mathrm{U}(3)$ nonet.
The explicit form of $\Phi$ is given by
\begin{equation}
\label{eq:PHI}
\Phi\equiv\sqrt{2}\,\mathrm{diag}\left(
\frac{1}{\sqrt{N_f}}\frac{\eta'}{F_{\eta'}}+\frac{1}{\sqrt{6}}\frac{\eta}{F_\eta},\,\,
\frac{1}{\sqrt{N_f}}\frac{\eta'}{F_{\eta'}}+\frac{1}{\sqrt{6}}\frac{\eta}{F_\eta},\,\,
\frac{1}{\sqrt{N_f}}\frac{\eta'}{F_{\eta'}}-\frac{2}{\sqrt{6}}\frac{\eta}{F_\eta}
\right)=\mathrm{diag}\left(\Phi_{uu},\Phi_{dd},\Phi_{ss}\right).
\end{equation}
With the representation of Eq.~(\ref{eq:PHI}) where the pion weak-decay constant is normalized in $F_\pi\approx93$ MeV~\cite{Halperin:1998gx}.

If the interaction strengths among the instantons and anti-instantons are the same then
the vacuum expectation value (VEV)of the pure Yang-Millers topological charge density operator $Q^\mathrm{YM}_t$ satisfies the following relation:
\begin{equation}
\label{eq:YMQ}
\int_x\langle (Q^\mathrm{YM}_t)^2\rangle\equiv\chi^\mathrm{YM}_t=\frac{N}{V}.
\end{equation}
As noted above, in the instanton configuration, this quantity relates to the instanton number density $N/V$ and can be defined as the topological susceptibility of the pure YM action $\chi^\mathrm{YM}_t.$~\cite{Diakonov:1997sj}.
Performing the trace and determinant over the flavor indices in Eq.~(\ref{eq:TCO}) with the following identities:
\begin{eqnarray}
\label{eq:ID}
\frac{1+\gamma_5}{2}\exp[i\Phi]\pm\frac{1-\gamma_5}{2}\exp[-i\Phi]
&=&(\gamma_5)^{\frac{1\mp1}{2}}\exp[i\gamma_5\Phi],
\cr
\left(\mathrm{det}_{f}\left[e^{i\Phi}\right] \right)^{\frac{1}{N_f-1}}\pm\left(\mathrm{det}_{f}\left[e^{-i\Phi}\right]\right)^{\frac{1}{N_f-1}}&=&\exp\left[\frac{i\mathrm{tr}_f\Phi}{N_f-1}\right]\pm\exp\left[\frac{-i\mathrm{tr}_f\Phi}{N_f-1}\right]=
2\left(
\begin{array}{c}
\cos\\i\sin
\end{array}
\right)i\sin\left[\frac{\sqrt{2N_f}\eta'}{F_{\eta'}(N_f-1)}\right],
\end{eqnarray}
we finally arrive at a rather concise expression for $Q_t(x)$:
\begin{eqnarray}
\label{eq:TCO1}
Q_t(x)&=&\int_k\int_pe^{i(k-p)\cdot x}
\sqrt{M_q(k)M_q(p)}\sum_{q}q^\dagger(k)
\left(\gamma_5+\delta I_{4\times4} \right)\exp[i\gamma_5\Phi_{qq}]q(p)
\cr&-&2(N_f-1)\left(\frac{\sqrt[3]{M_uM_dM_s}N_c}{\lambda^{1/N_f}}\right)^{\frac{N_f}{N_f-1}}
\left(\sin\left[\frac{\sqrt{2N_f}\eta'(x)}{F_{\eta'}(N_f-1)}\right]
-i\delta\cos\left[\frac{\sqrt{2N_f}\eta'(x)}{F_{\eta'}(N_f-1)}\right]\right)
\cr
&=&\int_k\int_pe^{i(k-p)\cdot x}
\sqrt{M_q(k)M_q(p)}\sum_{q}q^\dagger(k)\gamma_5(1+i\gamma_5\Phi_{qq})q(p)
-\chi^\mathrm{YM}_t\left(\frac{\sqrt{2N_f}\eta'(x)}{F_{\eta'}}-i(N_f-1)\delta\right)+\cdots
\end{eqnarray}
where we set $M_q(k)\equiv M_qF(k)$.
From the first line to the second of  Eq.~(\ref{eq:TCO1}), we have used the small-field approximation: $\eta'/F_{\eta'}\ll1$.

Taking Eqs.~(\ref{eq:TS}) and (\ref{eq:TCO1}) into account, the leading order diagrams $\sim\mathcal{O}(N_c)$ for the topological susceptibility $\chi_t$ in NL$\chi$QM are shown in Fig.~\ref{FIG0}. The solid and dash line indicates the quark and PS meson, whereas the solid ($\blacksquare$) and blank ($\square$) squares represent the vertices in $\mathcal{O}(1/\sqrt{N_c})$ and $\mathcal{O}(\sqrt{N_c})$, respectively, and the solid circle ($\bullet$) stands for the vertex in $\mathcal{O}(1)$. After straightforward calculations of the diagrams in according to $Q_t(x)$ in Eq.~(\ref{eq:TCO1}), the explicit expressions of each contributions for $\chi_t$ are given as follows:
\begin{eqnarray}
\label{eq:TST}
\chi_t&=&\left(\chi^a_t+\chi^b_t+\chi^c_t\right)_{q}+\chi^\phi_t+\chi^\Delta_t
=\chi^q_t+\chi^\phi_t+\chi^\Delta_t,
\cr
\chi^a_t&=&2L+L_s,\,\,\,\,
\chi^{b}_t=-\frac{2}{3}\left[\frac{(2L+L_s)^2}{F^2_{\eta'}m_{\eta'}^2}+\frac{2(L-L_s)^2}{F^2_{\eta}m_{\eta}^2}\right],\,\,\,\,
\chi^{c}_t=-4\left[\chi^\mathrm{YM}_t\frac{(2L+L_s)}{F^2_{\eta'}m_{\eta'}^2}\right],
\cr
\chi^\phi_t&=&\left(\chi^\mathrm{YM}_t\right)^2\frac{2N_f}{F^2_{\eta'}m^2_{\eta'}},
\,\,\,\,
\chi^\Delta_t=\delta^2(N_f-1)^2\left(\chi^\mathrm{YM}_t\right)^2
=\left(\chi^\mathrm{YM}_t\right)^3(N_f-1)^2\left(\sum_q\frac{1}{m_q\Sigma_q}\right)^2,
\end{eqnarray}
with the following definitions:
\begin{equation}
\label{eq:LFUNC}
L_{q}\equiv4N_c\int_k\frac{M^2_q(k)}{k^2+M^2_q(k)},\,\,\,L\equiv L_u=L_d,\,\,\,\,
L_0=4N_c\int_k\frac{M^2_0(k)}{k^2+M^2_0(k)}=\frac{N}{V}=\chi^\mathrm{YM}_t.
\end{equation}
Note that $\chi^\Delta_t$ survives only in the chiral limit $m_q\sim0$ because that $\delta\ll1$.

It is interesting to examine our result of Eq.~(18) in the quenched limit (QL) and the chiral limit.
We know that in the quenched limit $\chi^\mathrm{YM}_t\approx\chi^\mathrm{QL}_t$ and $\chi^q_t=0$, since the quark loop contributions are all suppressed in the quenched limit.
Moreover $\chi^\Delta_t$ is also suppressed in the large $N_{c}$ limit, according to $\delta\sim \mathcal{O}(1/N_{c})$. Hence, we can write the followings:
\begin{equation}
\label{eq:QL}
\chi^\mathrm{QL}_t=\left(\chi^\mathrm{YM}_t\right)^2
\frac{2N_f}{F^2_{\eta'}m^2_{\eta'}}\approx\left(\chi^\mathrm{QL}_t\right)^2
\frac{2N_f}{F^2_{\eta'}m^2_{\eta'}}\Rightarrow
\chi^\mathrm{QL}_t\approx\frac{F^2_{\eta'}m^2_{\eta'}}{2N_f}.
\end{equation}
Obviously, as shown in Eq.~(\ref{eq:QL}), we can obtain the Witten-Veneziano formula for the topological susceptibility in the quenched limit~\cite{Diakonov:1995qy}.

As for the chiral limit, we choose the conditions that $\chi^\mathrm{YM}_t\approx \chi^\mathrm{\chi L}_t$ and $\chi^q_t\ll\chi^\mathrm{\chi L}_t$, since $\chi^\Delta_t$ dominates in this limit $m_q\sim0$:
\begin{equation}
\label{eq:CL}
\chi^\mathrm{\chi L}_t=
\left(\chi^\mathrm{YM}_t\right)^3(N_f-1)^2\left(\sum_q\frac{1}{m_q\Sigma_q}\right)^2
\approx
\left(\chi^\mathrm{\chi L}_t\right)^3N_f^2\left(\sum_q\frac{1}{m_q\Sigma_q}\right)^2
\Rightarrow
\chi^\mathrm{\chi L}_t\approx\frac{1}{N_f}\left(\sum_q\frac{1}{m_q\Sigma_q}\right)^{-1},
\end{equation}
which is nothing but the Leutwyler-Smilga formula for the topological susceptibility in the chiral limit~\cite{Leutwyler:1992yt,DiGiacomo:1997wf}. Thus, as for $\chi_t$ with the physical current quark masses, i.e. physical limit (PL) with the conditions that $m_q>0$ and $N_f=3$, we can write
\begin{equation}
\label{eq:PL}
\chi^\mathrm{PL}_t=\chi^q_t+\chi^\phi_t.
\end{equation}
In the following Section, we represent the numerical result of the topological susceptibility $\chi_t=\chi^\mathrm{PL}_t$.

\section{Numerical results and discussions}
In this Section, we present our numerical results and the relevant discussions. The first task is to determine the value of the quark mass: $M_q=M_0+m_q$. It is the sum of the current quark mass $m_q$ and the constituent quark mass $M_0$ coming from S$\chi$SB. The value of $M_0$ can be determined in the chiral limit in LIM using the self-consistent (gap) equation Eq.~(\ref{eq:NOV}). By comparing Eq.~(\ref{eq:NOV}) with Eq.~(\ref{eq:LFUNC}), we identify that $N/V=L_0$ with $m_q=0$.
The form factor $F(k)$ comes from the Fourier transform of the quark zero-mode solution in the instanton vacuum. It consists of the combination of modified Bessel functions~\cite{Diakonov:2002fq}:
\begin{equation}
\label{eq:FF}
F(k)=2t\left[I_0(t)K_1(t)-I_1(t)K_0(t)-\frac{I_1(t)K_1(t)}{t}\right]_{t=\frac{|k|\bar{\rho}}{2}}.
\end{equation}
Here we choose the {\it standard} values for the LIM parameters: The (anti)instanton number density $N/V=1/\bar{R}^4=(1\,\mathrm{fm})^{-4}\approx(197\,\mathrm{MeV})^4$ and the average (anti)instanton size $1/\bar{\rho}=(1/3\,\mathrm{fm})^{-1}\approx591$ MeV~\cite{Diakonov:2002fq}. $M_0$ is found to be $341$ MeV
in according to the self-consistent equation in Eq.~(\ref{eq:NOV}).

The weak decay constants of $\eta$ and $\eta'$ mesons, $F_\eta$ and $F_{\eta'}$
are also needed in the computation of $\chi_t$ as shown in Eq.~(\ref{eq:TST}).
Their analytical expressions derived from the axial-vector current conservation read as~\cite{Nam:2007gf}:
\begin{equation}
\label{eq:FPI}
F^2_{q\bar{q}}=4N_c\int_k\frac{M^2_{q\bar{q}}(k)-\frac{k}{2}M_{q\bar{q}}(k)\frac{\partial M_{q\bar{q}}(k)}{\partial k}+\frac{k^2}{4}\left[\frac{\partial M_{q\bar{q}}(k)}{\partial k}\right]^2}{[k^2+M^2_{q\bar{q}}(k)]^2},\,\,\,\,M_{q\bar{q}}(k)\equiv\frac{M_q(k)+M_{\bar{q}}(k)}{2}
=\frac{(M_{q}+M_{\bar{q}})}{2}F^2(k).
\end{equation}
Here, $F_{q\bar{q}}$ stands for the weak-decay constant for a pseudo-scalar meson consisting of $q$ and $\bar{q}$. For instance, we have $(q,\bar{q})=(u,\bar{d})$ for $F_{\pi^+}$.  First, we compute $F_\pi$ and $F_K$ to check the validity of the model. Choosing $M_{u\bar{d}}=(M_u+M_{\bar{d}})/2\equiv M=346$ MeV for the pion and $M_{u\bar{s}}=(M+M_{\bar{s}})/2=(346+476)/2=411$ MeV for the kaon, we have that
\begin{eqnarray}
\label{eq:WDC}
\mathrm{NL}\chi\mathrm{QM}&:&F_\pi=91.01\,\mathrm{MeV},\,\,\,\,F_K=100.08\,\mathrm{MeV},
\cr
\mathrm{Empirical\,\,value}&:&F_\pi=93\,\mathrm{MeV},\hspace{0.6cm}F_K=108\,\mathrm{MeV},
\end{eqnarray}
where we have used $m_{u,d}\equiv m=5$ MeV and $m_s=135$ MeV.
Similarly, one obtains $F_{u\bar{u}}=F_{d\bar{d}}=91.01$ MeV and $F_{s\bar{s}}=108.28$ MeV in according to Eq.~(\ref{eq:FPI}).
Furthermore, the empirical value of $F_\eta$ is $(91.9\pm0.07)$ MeV from the PDG value~\cite{Agashe:2014kda}.
Hence we have
\begin{eqnarray}
\label{eq:WDC!}
\mathrm{NL}\chi\mathrm{QM}&:&F_{\eta}=96.77\,\mathrm{MeV},\,\,\,\,F_{\eta'}=102.53\,\mathrm{MeV}.
\cr
\mathrm{Empirical\,\,value}&:&F_{\eta}=(91.9\pm0.07)\,\mathrm{MeV}\,\,\,(F_{\eta'}=F_{\eta}\,\,\mathrm{for}\,\,N_c\to\infty).
\end{eqnarray}

The singlet and octet contributions of $F_{\eta}$ and $F_{\eta'}$ were investigated
by using the $\mathrm{U_A}(1)$ PCAC
in Ref.~\cite{Shore:2006mm}
Their results are $(F_{0\eta},F_{8\eta})=(22.8\pm5.7,98.4\pm1.4)$ MeV and $(F_{0\eta'},F_{8\eta'})=(104.2\pm4.0,-36.1\pm1.2)$ MeV.
Moreover, AdS/QCD calculation gives $(F_{0\eta},F_{8\eta})=(17.0,103.0)$ MeV and $(F_{0\eta'},F_{8\eta'})=(129.0,-35.1)$ MeV~\cite{Katz:2007tf}.
We list various estimations for $(F_{\eta},F_{\eta'})$ in Table~\ref{TABLE1}.
\begin{table}[h]
\begin{tabular}{|c||c|c|c|c|}
\hline
&NL$\chi$QM&$\mathrm{U_A}(1)$ PCAP~\cite{Shore:2006mm}&AdS/QCD~\cite{Katz:2007tf}&PDG~\cite{Agashe:2014kda}\\
\hline
$F_{\eta}$&$96.77$&$(22.8\pm5.7,98.4\pm1.4)$&$(17.0,103.0)$&$91.9\pm0.07$\\
\hline
$F_{\eta'}$&$102.53$&$(104.2\pm4.0,-36.1\pm1.2)$&$(129.0,-35.1)$&$-$\\
\hline
\end{tabular}
\caption{The values for $F_{\eta}$ and $F_{\eta'}$ [MeV]. As for Refs.~\cite{Shore:2006mm,Katz:2007tf}, we represent them in the form of $(F_{0h},F_{8h})$, where $h=(\eta,\eta')$.}
\label{TABLE1}
\end{table}
Our value of $F_{\eta'}$ is close to the values of $F_{0\eta'}$ of other calculations and
our value of $F_{\eta}$ is close to the values of $F_{8\eta}$ of other calculations.
Since we know that $F_{\eta'}\sim F_{0\eta'}$ and  $F_{\eta}\sim F_{8\eta}$, i.e. the mixing angle between $\eta$ and $\eta'$ is expected to be large,
such that we will use our values of $F_{\eta}$ and $F_{\eta'}$ to compute $\chi_t$ within NL$\chi$QM.

We make further test in NL$\chi$QM by computing the quark condensate, $\Sigma_q\equiv i\langle q^\dagger q\rangle$. Note that $\Sigma_q$ also appears in the LS formula for $\chi_t$ in Eq.~(\ref{eq:CL}). Within NL$\chi$QM, the chiral condensate is given by~\cite{Nam:2007gf}:
\begin{equation}
\label{eq:QC}
\Sigma_q=4N_c\int_k\left[\frac{M_q(k)}{k^2+M^2_q(k)}-\frac{m_q(k)}{k^2+m^2_q(k)}\right].
\end{equation}
Here, we set $m_q(k)=m_q F^2(k)$ to tame the UV divergence. As for $q=(u\sim d,s)$, we have the following values in NL$\chi$QM:
\begin{equation}
\label{eq:QCV}
\Sigma_{u,d}\equiv \Sigma=(251.31\,\mathrm{MeV})^3,
\,\,\,\,\,\,\,\,
\Sigma_s=(242.93\,\mathrm{MeV})^3.
\end{equation}
The ratio of those values $R=\Sigma/\Sigma_s$ is about $0.91$ locating inside other empirical and theoretical estimations $(0.75\sim1.05)$~\cite{Nam:2006ng}. The PS-meson masses can be expressed by the combination of the quark condensate, current quark mass, and pion weak decay constant via $\chi$PT in the large $N_c$ limit as follows~\cite{Evans:1996kf}:
\begin{equation}
\label{eq:GOR}
m^2_\pi =\frac{2m\Sigma}{F^2_\pi},\,\,\,\,
m^2_K =\frac{m\Sigma+m_s\Sigma_s}{F^2_\pi},\,\,\,\,
m^2_\eta=\frac{2}{3}\frac{(m\Sigma+2m_s\Sigma_s)}{ F^2_\pi}.
\end{equation}
Using the numerical values of the condensates and weak-decay constants within the present model as shown above, we obtain the followings:
\begin{eqnarray}
\label{eq:MASS}
\mathrm{NL}\chi\mathrm{QM}&:&
m_\pi=138.43\,\mathrm{MeV},\,\,\,
m_K=493.16\,\mathrm{MeV},\,\,\,
m_\eta=563.82\,\mathrm{MeV},\,\,\,
m_{\eta'}=954.76\,\mathrm{MeV},
\cr
\mathrm{Experiment}&:&
m_\pi=138.04\,\mathrm{MeV},\,\,\,
m_K=495.67\,\mathrm{MeV},\,\,\,
m_\eta=547.86\,\mathrm{MeV},\,\,\,
m_{\eta'}=957.78\,\mathrm{MeV}.
\end{eqnarray}
Note that the value of $m_{\eta'}$ is determined by equating Eqs.~(\ref{eq:CHPT}) and (\ref{eq:TST}).
From Eq.~(\ref{eq:TST}) the value of $\chi_{t}$ depends on the value of $m_{\eta'}$, however, the values of $\chi_t$ and
$m_{\eta'}$ is also related by  Eqs.~(\ref{eq:CHPT}) such that we can determine $m_{\eta'}$ in NL$\chi$QM. It is worth mentioning that various quantities, such as the PS-meson mass, the weak decay constant, and the quark condensate, are all reproduced well within the model with only two free parameters ($\bar{\rho}$ and $\bar{R}$). All the relevant parameters and inputs of our numerical calculation of the topological susceptibility are summarized in Table~\ref{TABLE2}.
\begin{table}[h]
\begin{tabular}{|c|c|c|c|c|c|c|c|c|c|c|}
\hline
$1/\bar{\rho}$
&$\sqrt[4]{N/V}$
&$\hspace{0.4cm}M_0\hspace{0.4cm}$
&$m_{u,d}\equiv m$
&$\hspace{0.4cm}m_s\hspace{0.4cm}$
&$M_{u,d}\equiv M$
&$\hspace{0.4cm}M_s\hspace{0.4cm}$
&$\hspace{0.4cm}m_\eta\hspace{0.4cm}$
&$\hspace{0.4cm}m_{\eta'}\hspace{0.4cm}$
&$\hspace{0.4cm}F_\eta\hspace{0.4cm}$
&$\hspace{0.4cm}F_{\eta'}\hspace{0.4cm}$\\
\hline
$591$&$197$&$341$&$5$&$135$&$346$
&$476$&$563.82$&$954.76$&$96.77$&$102.53$\\
\hline
\end{tabular}
\caption{Model parameters and theoretical inputs [MeV] in the present calculations.}
\label{TABLE2}
\end{table}

Using Eq.~(\ref{eq:TST}) and all the parameters discussed above, we present our main results for the full, Witten-Veneziano (WV), and Leutwyler-Smilga (LS) formulae with physical input values as follows:
\begin{equation}
\label{eq:FTS}
\chi_{t}=(165.57\,\mathrm{MeV})^4,
\,\,\,\,\chi^\mathrm{WV}_t=(194.30\,\mathrm{MeV})^4,
\,\,\,\,\chi^\mathrm{LS}_t=(162.54\,\mathrm{MeV})^4.
\end{equation}
$\chi^\mathrm{WV}_t=\chi^\mathrm{QL}_t$ since the WV formula is derived in the large $N_{c}$ limit and the dynamical quark contribution is neglected. Although we are able to make our values of $\chi_t$ to be closer to the empirical one by adjusting the free parameters, we prefer to not to do so since it does not tell much for relevant physics. Our estimation is about $10\%$ smaller than the empirical value $\chi_t=(175\pm5\,\mathrm{MeV})^4$. Furthermore, we observe that $\chi_{t}<\chi^\mathrm{QL}_t$ in general in according to Eq.~(\ref{eq:FTS}). It means that the dynamical quark contributions reduce the topological susceptibility by $(10\sim20)\,\%$ within NL$\chi$QM.

Finally, we would like to make comparison between our results and other theoretical estimations on $\chi_t$. In Ref.~\cite{Fukushima:2001hr},  $\chi_t$
was computed by using the NJL model with the standard parameter sets with the additional ring diagrams contribution in terms of the mean-field approach.
They found that $\chi_t=(166\,\mathrm{MeV})^4$ with $m_\eta=487$ MeV. This value is very close to our full and LS values.
Furthermore, this result can be improved by taking into account that the standard parameter set for determining $m_{\eta'}$ is not suitable
for computing $\chi_t$ due to the lack of confinement in the NJL model.

There are many lattice computations of the topological susceptibility, and it is interesting to compare those results with ours.
The LQCD simulation of the CP-PACS collaboration~\cite{AliKhan:1999zi,AliKhan:1999zp} gives $m_\eta\approx863$ MeV
and $\chi_t=(178(9)\,\mathrm{MeV})^4$
in the continuum limit.
This simulation is the two-flavor full QCD configuration with the improved actions, i.e. a tree-level Symanzik-improved and the RG improved Iwasaki actions.
The UKQCD collaboration performed the LQCD simulations, in which the color $\mathrm{SU}(3)$ Wilson-gauge action is coupled to clover improved Wilson fermions. They computed $\chi_t$ in $N_f=2$~\cite{Hart:2000wr,Hart:2000gh}, and obtained $\chi_t=(177\,\mathrm{MeV})^4$.
Those results are substantially larger than ours, however, it may relate to the fact that their dynamical quarks are only two flavors.

The LQCD simulation by the SESAM and T$\chi$L collaborations with two flavors of dynamical Wilson fermions have also been carried out. They gave $\chi_t\approx(183\,\mathrm{MeV})^4$ for a certain lattice set, whereas $\chi^\mathrm{QL}_t\approx(199\,\mathrm{MeV})^4$ for the quenched case~\cite{Bali:2001gk}. Note that the hierarchy between the full and the quenched results is similar to our observation $\chi_{t}<\chi^\mathrm{QL}_t$. In Ref.~\cite{Grunberg:1984pa}, using the QCD sum rue (QCDSR) with the help of the flavor $\mathrm{SU}(2)$ LQCD data for the gluon condensate, it gave $\chi_t\approx(180\,\mathrm{MeV})^4$ employing the renormalization scale $\Lambda_{\overline{\mathrm{MS}}}=175$ MeV.  In Ref.~\cite{Chiu:2008jq}, via LQCD simulation using $(2+1)$-flavors of domain-wall fermions, the authors estimate $\chi_t=(153\sim187\,\mathrm{MeV})^4$, depending on the quark mass. The comparison between our result and the other theoretical estimations are listed in Table~\ref{TABLE4}.
\begin{table}[h]
\begin{tabular}{|c||c|c|c|c|c|c|c|c|c|}
\hline
&NL$\chi$QM&NJL~\cite{Fukushima:2001hr}&LQCD~\cite{DelDebbio:2004ns}&LQCD~\cite{Hart:2000wr,Hart:2000gh}&LQCD~\cite{Bali:2001gk}&LQCD~\cite{Aoki:2000kp}&LQCD~\cite{AliKhan:1999zi,AliKhan:1999zp}&LQCD~\cite{Chiu:2008jq}&QCDSR~\cite{Grunberg:1984pa}\\
\hline
$\sqrt[4]{\chi_t}$&$165.57\,(194.30_\mathrm{QL})$&$166$&$191\pm5$&$177$
&$183\,(199_\mathrm{QL})$&$174$&$178(9)$&$153\sim187$&$180$\\
\hline
\end{tabular}
\caption{Topological susceptibility $\sqrt[4]{\chi_t}$ [MeV] from various theoretical estimations.}
\label{TABLE4}
\end{table}

From all the comparisons, our main results are comparable to the $N_f=2$ and $(2+1)$ LQCD data, showing some deviations. As expected, approaches based on QCD, such as the NJL model and QCDSR, provide similar values to ours with below $10\%$ differences.

\section{Conclusion and outlooks}
We have studied the QCD topological susceptibility $\chi_t$ by employing the nonlocal chiral quark model (NL$\chi$QM). This model is based on the liquid-instanton model of the QCD-vacuum configuration. Various nonperturbative quantities, such as the PS-meson masses, the weak decay constants, and the quark condensates, are well reproduced within this model with only two free parameters ($\bar{\rho}$ and $\bar{R}$) in principle. We then derived a concise expression of the topological charge-density operator from NL$\chi$QM with the determinant manifesting the $\mathrm{U_A}(1)$ anomaly. The value of $\chi_t$ was computed by using the correlation functions of the topological-charge density operator. 

We find that such an effective model gives correct expressions for the Witten-Veneziano (WV) and Leutwyler-Smilga (LS) formulae for $\chi_t$. Moreover, The {\it ring} diagrams appearing in the generic NJL model in terms of the mean-field approach, are naturally emerged as the PS-meson propagations in NL$\chi$QM. Furthermore, we obtain  $\chi_{t}=(165.57\,\mathrm{MeV})^4$, $\chi^\mathrm{WV}_t=(194.30\,\mathrm{MeV})^4$, and $\chi^\mathrm{LS}_t=(162.54\,\mathrm{MeV})^4$, which are comparable with the other theoretical estimations. Note that, phenomenologically, it is $\chi_{t}=(175\pm5\,\mathrm{MeV})^4$. We also figure out that there is a tendency that $\chi_{t}<\chi^\mathrm{QL}_t$, showing the $(10\sim20)\,\%$ decrease with the dynamical-quark contributions in comparison to the QL cases. A similar tendency is also found in a full-LQCD simulation.

We note that topological susceptibility  $\chi_t$ is deeply related to the axion potential as a function of temperature ($T$)~\cite{Borsanyi:2015cka}. Moreover, $\chi_t$ reveals rich information on the QCD phase diagram as an order parameter of $\mathrm{U_A}(1)$ symmetry and its relation with
S$\chi$SB is also studied~\cite{Jiang:2015xqz}. As already explored in our previous works~\cite{Nam:2011vn,Nam:2013fpa}, the present model has provided successful descriptions for various QCD properties at finite $T$ as well as quark chemical potential $\mu_q$. It is based on the instanton QCD-vacuum configuration which is responsible for the nontrivial mechanisms of S$\chi$SB and its (partial) restoration. Hence, we expect that the careful studies on $\chi_t$ in medium within the current theoretical framework will shed light on understanding of the strongly-interacting systems at finite $T$ and/or $\mu_q$. Related works are under progress and will appear elsewhere.
\section*{Acknowledgements}
S.~i.~N. is grateful to the hospitality for his stay at National Taiwan University (NTU), where the present work was performed partially, with the financial supports from NSC102-2112-M-033-005-MY3 from Ministry of Science and Technology (MOST) of Taiwan. His work is also in part supported by the NRF grant funded by MEST (Center for Korean J-PARC Users, Grant No. NRF-2013K1A3A7A06056592). He also appreciates the fruitful discussions with K.~H.~Woo.
C.~W.~K. is supported by NSC102-2112-M-033-005-MY3 from Ministry of Science and Technology (MOST) of Taiwan.

\end{document}